\documentclass[12pt]{article}
\begin{document}
\title{Oscillator Models for the Universe and Dark Energy}
\author{B.G. Sidharth\\
Centre for Applicable Mathematics \& Computer Sciences\\
B.M. Birla Science Centre, Adarsh Nagar, Hyderabad - 500 063 (India)}
\date{}
\maketitle
\begin{abstract}
We consider two situatiions that can be modeled by oscillators. First we show that the universe itself is a normal mode of Planck scale oscillators. Next we show that a background Zero Point Field or dark energy can be modeled by electromagnetic oscillators. This in turn leads to an observable ``fluctuational'' current in a solenoid.
\end{abstract}
\section{Planck Scale Oscillators}
In an earlier communication \cite{r1}, it was argued that a typical
elementary particle like a pion could be considered to be the result of
$10^{40}$ evanescent Planck scale particles. The argument was based on 
random motions and also on the modification to the Uncertainity Principle.
We will now consider the problem from a totally different point of view,
which not only reconfirms the above result, but also enables an elegant
extension to the case of the entire universe itself.\\
We first consider an array of $N$ particles, spaced a distance $\Delta x$
apart, which behave like oscillators, that is as if they were connected by
springs. We then have \cite{r2,r3}
\begin{equation}
r  = \sqrt{N \Delta x^2}\label{e1}
\end{equation}
\begin{equation}
ka^2 \equiv k \Delta x^2 = \frac{1}{2}  k_B T\label{e2}
\end{equation}
where $k_B$ is the Boltzmann constant, $T$ the temperature, $r$ the extent  and $k$ is the 
spring constant given by
\begin{equation}
\omega_0^2 = \frac{k}{m}\label{e3}
\end{equation}
\begin{equation}
\omega = \left(\frac{k}{m}a^2\right)^{\frac{1}{2}} \frac{1}{r} = \omega_0
\frac{a}{r}\label{e4}
\end{equation}
We now identify the particles with Planck masses, set $\Delta x \equiv a = 
l_P$, the Planck length. It may be immediately observed that use of 
(\ref{e3}) and (\ref{e2}) gives $k_B T \sim m_P c^2$, which ofcourse agrees 
with the temperature of a black hole of Planck mass. Indeed 
Rosen had shown that a Planck mass particle at the Planck scale  can be considered to be a
universe in itself \cite{r4}.\\
We now use the fact alluded to that  a typical elementary particle
like the pion can be considered to be the result of $N \sim 10^{40}$ Planck
masses (Cf.ref.\cite{r1}). Using this in (\ref{e1}), we get $r \sim l$, the pion
Compton wavelength as required. Further, using (\ref{e2}), (\ref{e3}) and
(\ref{e4}), we get for a pion, 
$$k_ B T = \frac{m^3 c^4 l^2}{\hbar} = mc^2,$$
which ofcourse is the well known formula for the Hagedorn temperature for
elementary particles like pions. In other words, this confirms the conclusions
in \cite{r1}, that we can treat an elementary particle as a series of some
$10^{40}$ Planck mass oscillators.\\
However it must be observed from 
(\ref{e2}) and (\ref{e3}), that while the Planck mass gives the highest
energy state, an elementary particle like the pion is in the lowest energy
state. This explains why we encounter elementary particles, rather than
Planck mass particles in nature. Infact as already noted \cite{r5,r6}, a Planck
mass particle decays via the Bekenstein radiation within a Planck time
$\sim 10^{-42}secs$. On the other hand, the lifetime of an elementary particle
would be very much higher.\\
Using the fact that the universe consists of $n \sim 10^{80}$ elementary
particles like the pions, the question is, can we think of the universe as
a collection of $10^{120}$ Planck mass oscillators? This is what we will now
show. Infact if we use equation (\ref{e1}) with
$$N \sim 10^{120},$$
we can see that the extent $r \sim 10^{28}cms$ which is of the order of the diametre of the
universe itself. Next using (\ref{e4}) we get
\begin{equation}
n \hbar \omega_0^{(min)} \frac{l_P}{10^{28}} \approx m_P c^2 \times 10^{60}\label{e5}
\end{equation}
which gives the correct mass of the universe. In other words the universe
itself can be considered to be a normal mode of Planck scale oscillators.
\section{Oscillators and Dark Energy}
The existence of the Zero Point Field (ZPF) was realised even by Max Planck. 
In its later version it has come to be known as Quantum Vaccuum, and
lately it has been possible to identify it with the mysterious dark
energy. The Zero Point Field arises because of the fact that though
classically, an oscillator in the ground state has zero energy, Quantum
Mechanically, owing to the uncertainity principle, there is an energy
fluctuation about the zero energy level.\\
Interestingly there have been two schools of thought. Zero Point Field, 
according to Quantum theoriests is a secondary effect arising from the
already present oscilaltors. However according to what has come to be
known as Stochastic Electrodynamics, the ZPF is primary, and infact
Quantum Mechanics can be deduced therefrom.  This is a chicken and egg
situation.\\
Recently the ZPF has come into focus once again because of the 
observed ever expanding, accelerating feature of the universe. In other
words there is a large scale repulsion represented by a cosmological constant 
$\Lambda$ once invented and then
rejected by Einstein, and this cosmic repulsion could be attributed to a
mysterious all pervading dark energy, which can be identified with the
ZPF: this could be the mechanism which drives the cosmic
expansion and acceleration (Cf.ref.\cite{r1a}) for a 
review).\\
Two of the earliest realisations of the ZPF were in the form of the Lamb shift
and the Casimir effect.\\
In the case of the Lamb shift, as is well known, the motion of an orbiting
electron gets affected by the background ZPF. Effectively there is an
additioinal field, over and above that of the nucleus. This additional
potential, as is well known is given by
 \cite{r2a}
$$\Delta V (\vec r) = \frac{1}{2} \langle (\Delta r)^2 \rangle \nabla^2 V
(\vec r)$$
The additional energy
$$\Delta E = \langle \Delta V (\vec r) \rangle$$
contributes to the observed Lamb shift which is $\sim 1000 mc/sec$.\\
The essential idea of the Casimir effect is that the interaction between
the ZPF and matter leads to macroscopic consequences. For example if we
consider two parallel metallic plates in a conducting box, then we should
have a Casimir force given by
 \cite{r3a}
$$F = \frac{-\pi^2}{240} \frac{\hbar cA}{l^4}$$
where $A$ is the area of the plates and $l$ is the distance between them.
More generally, the Casimir force is a result of the
boundedness or deviation from a Euclidean topology of or in the Quantum
Vaccuum. These Casimir forces have been experimentally demonstrated
\cite{r4a,r5a,r6a,r7a}.\\
Returning to the ZPF
as the ubiquitous dark energy, we observe that \cite{r8a}, a fluctuating electromagnetic
field can be modelled as an infinite collection of independent harmonic oscillators.
Quantum Mechanically, the ground state of the harmonic oscillators is described by
$$\psi (x) = \left(\frac{m\omega}{\pi \hbar}\right)^{1/4} 
e^{-(m\omega/2\hbar)x^2}$$
which exhibits the probability for the oscillator to fluctuate, mostly in the region
given by
$$\Delta x \sim (\hbar /m\omega )^{1/2}$$
An infinite collection of such oscillators can be modelled by
$$\psi (\xi_1,\xi_2,\cdots ) = const. \exp [-(\xi^2_1 + \xi^2_2 + \cdots )],$$
which gives the probability amplitude for an electromagnetic field configuration
$B(x,y,z), \xi_1$, etc. being the Fourier coefficients. Finally, as a consequence there
is a fluctuating magnetic field given by
\begin{equation}
B = \frac{\sqrt{\hbar c}}{l^2}\label{eD}
\end{equation}
where $l$ is the extent over which the fluctuation is measured. Further
these fluctuations typically take place within the time $\tau$, a typical
elementary particle Compton time (Cf.ref.[1]). This begs the question whether such
ubiquotous fields could be tapped for terrestrial applications or otherwise.\\
We now invoke the well known result from macroscopic physics that the
current in a coil is given by
\begin{equation}
\imath = \frac{NBA}{R\Delta t}\label{eE}
\end{equation}
where $N$ is the numer of turns of the coil, $A$ is its area and $R$ the
resistance.\\
Introducing (\ref{eD}) into (\ref{eE}) we deduce that a coil in the ZPF
would have a fluctuating electric current given by
\begin{equation}
\imath \approx \frac{NA}{R} \cdot \frac{e}{l^2\tau}\label{eF}
\end{equation}
In principle it should be possible to harness the current (\ref{eF}). While this current is small, if we have a superconductor, then $R$ would be very small and the current would be much larger. The question is, whether such an application is possible, on the earth or in an orbiting space craft, for example.\\
Finally, it may be remarked that the usual ZPF of Quantum Theory leads to an unduly large cosmological constant \cite{r9a}. However, in the model described in \cite{r5,r1a}, we get the observed small cosmological constant.


\begin{thebibliography}{99}
\bibitem {r1} B.G. Sidharth, Found. of Phys.Lett., 15 (6), December 2002, 577-583.
\bibitem {r2}Y. Jach NG and H. Van Dam, Mod.Phys.Lett.A., 9 (4), 1994, p.335-340.
\bibitem {r3} D.L. Goodstein, ``States of Matter'', Dover Publications Inc., New York, 1985, p.160ff.
\bibitem {r4} N. Rosen, Int.J.Th.Phys., 32 (8), 1993, p.1435-1440.
\bibitem {r5} B.G. Sidharth, ``Chaotic Universe: From the Planck to the Hubble Scale'', Nova Science Publishers Inc., New York, 2001, p.65ff.
\bibitem {r6} B.G. Sidharth, Chaos Solitons and Fractals, 12 (1), 2000, 173-178. 
\bibitem {r1a} B.G. Sidharth, Chaos, Solitons and Fractals, 16(4), 2003, p.613-620.
\bibitem {r2a} J.D. Bjorken, and S.D. Drell "Relativistic Quantum Mechanics", Mc-Graw
Hill, New York, 1964, p.39.
\bibitem {r3a} L. de la Pena and A.M. Cetto, Found.of Phys., Vol.12, No.10, 1982,
p.1017-1037 and references therein.
\bibitem {r4a} V.N. Mostepanenko and N.N. Trunov, Sov.Phys.Usp., Vol. 31, No.11, 1988,
p.965-987.
\bibitem {r5a} V.B. Svetovoy and M.V. Lokhanin, Mod.Phys.Lett.A., Vol.15, Nos.22 \& 23
2000, p.1437-1444.
\bibitem {r6a} P.K. Anastasovski, T.E. Bearden, et al., Found.of Phys.Lett., Vol.13, No.3,
2000, p.289-296.
\bibitem {r7a} A. Bulgac and A. Wirzba, Phys.Rev.Lett., Vol.87, No.12, 2001, p.120404-1-
120404-4.
\bibitem {r8a} C.W. Misner, K.S. Thorne and J.A. Wheeler, "Gravitation", W.H. Freeman,
San Francisco, 1973, pp.1180ff.
\bibitem {r9a} S. Weinberg, Phys.Rev.Lett., 43, 1979, p.1566.
\end{thebibliography}
\end{document}